\documentclass[bbm,amsfonts,amsmath,amssymb,12pt]{article} 
\usepackage{amssymb,amsmath} 
\usepackage{amscd} 
\usepackage{fancyhdr} 
\usepackage{color} 
\usepackage{hyperref} 
\usepackage{epsfig} 
\usepackage{graphicx}
\begin{document}
\begin{titlepage} 
 
\begin{flushright} 
\small 
\end{flushright} 
 
\vspace {1cm} 
 
 \begin{center}
\Large \bf Chiral Models in Noncommutative ${\cal N=}1/2$ \\Four Dimensional Superspace
 \end{center}
\vskip 1cm \centerline{\large Thomas A. Ryttov\footnote{E-mail: ryttov@nbi.dk} and Francesco Sannino\footnote{E-mail: sannino@nbi.dk}} \vskip 0.1cm 
\vskip .5cm 
\begin{center} 
The Niels Bohr Institute, Blegdamsvej 17, DK-2100 \\Copenhagen \O, Denmark
\end{center} 
\vskip 1cm 
 
\begin{abstract} 
We derive the component Lagrangian for a generic ${\cal N}=1/2$ supersymmetric chiral model with an arbitrary number of fields in four space-time dimensions. We then investigate a toy model in which the deformation parameter modifies the undeformed potential near the origin of the field space in a way which suggests possible physical applications.
\end{abstract}

\end{titlepage} 

\section{Introduction}
\label{sec:introduction}
Understanding the underlying structure of space and time poses continuous challenges. 
It is hence interesting to study possible deformations of ordinary space-time which are 
mathematically under control. Supersymmetric theories provide a natural laboratory to explore new 
ideas. Of particular current interest are the non(anti)commutative deformed versions of the ${\cal N} =1$ superspace \cite{Seiberg:2003yz} according to which one requires the Grassmanian variables of the 
ordinary superspace $\theta^{\alpha}$ to
satisfy the following anticommutator relation:
\begin{equation}
  \label{eq:70}
  \{\theta^{\alpha},\theta^{\beta}\}=C^{\alpha\beta} \ .
\end{equation}
Since the $\bar{\theta}$ variables are taken to anticommute a Clifford deformation \cite{Hirshfeld:2002ki} of the superspace must be performed in the Euclidean space. Here the $\theta$ and $\bar{\theta}$ are independent \cite{Klemm:2001yu}. This also implies that only half of the original supersymmetry is kept intact \cite{Seiberg:2003yz,Ferrara:2000mm}. Another interesting point is that these theories naturally 
emerge as an effective low energy description of certain superstring theories in constant graviphoton backgrounds \cite{Ooguri:2003tt}. This is however not an entirely new subject \cite{{Casalbuoni:1975hx},{Schwarz:1982pf}}.

Seiberg has shown that the noncommutative\footnote{Note that
we refer to the nonanticommutative space as noncommutative while the standard anticommutative space 
is abbreviated as commutative.} deformation of the Wess-Zumino model and of ${\cal N}=1$ super Yang-Mills theories results in the addition of a finite number of higher dimensional operators with respect to the 
undeformed Lagrangian \cite{Seiberg:2003yz}. The deformed theories preserve locality, Lorentz symmetry 
and are renormalizable \cite{Britto:2003aj}. The study of instantons \cite{Imaanpur:2003jj} in this 
framework has also been 
a subject of much interest as well as the lower dimensional cousin theories \cite{Ivanov:2003te,Chandrasekhar:2003uq,Chandrasekhar:2004ti,{Alvarez-Gaume:2005pj},{Chandrasekhar:2005hn}}. 
Another interesting application of the deformed superspace used to analyze spectral properties in supersymmetric gluondynamics and 
one-flavor QCD has been proposed in \cite{Gorsky:2004nb}.

Here we construct the component 
form of the most general chiral model theory with an arbitrary number of chiral superfields in the deformed superspace theory in four dimensions. This generalizes
the work in \cite{Azorkina:2005mx} and \cite{Hatanaka:2005rg} which focussed on a single chiral superfield. Due to the fact that the product of chiral superfields 
does not commute in the deformed superspace one must use an ordering procedure which renders the 
generalization to an arbitrary number of fields quite involved. 

We elucidate some issues related to how to handle the K\"{a}hler of the theory. More precisely we show that certain Lagrangian terms, present in \cite{Azorkina:2005mx}, are actually absent when considering the fully symmetrized deformed K\"{a}hler of the theory with respect to both chiral and antichiral superfields simultaneously.

We recall that 
supersymmetric chiral models with an arbitrary number of fields are relevant since 
they have been used, for example, to investigate the vacuum structure of supersymmetric gauge theories \cite{Intriligator:1995au}. Besides many models of spontaneously broken supersymmetry involve more than one field and/or nonperturbatively induced superpotentials. 
Having determined the general chiral model for $N=1/2$ with an arbitrary number of superfields allows 
immediately to investigate the deformations of a number of low energy superpotentials derived in the standard supersymmetric case.

We then propose an instructive toy model in which the potential of the theory at the component level, and after having integrated out the auxiliary field, 
contains an infinite number of terms. Interestingly the full series in the deformation parameter can be summed. Another amusing property of the 
final potential is that it is a function only of the product of the scalar and the antiscalar field. 
In the Euclidean space the antiscalar field is not merely the complex conjugated of the scalar field. Although we are forced, by the very 
definition of the deformed superspace, to be in the Euclidean space the structure of the potential is so intriguing to us that we have been 
tempted to analytically continue it in the Minkowski space. If the determinant of the deformation parameter is assumed real then 
also the potential is real and positive. One can show that the vacuum structure of the theory is unchanged with respect to the undeformed space while the 
potential is largely affected in the infrared, i.e. for small values of the bosonic field values. It naturally introduces a new maximum with a finite value of the potential while the theory still preserves Lorentz symmetry \cite{Seiberg:2003yz}. 

The presence of two scales, i.e. the one associated to the deformation parameter, which may be linked to the gravity scale, 
together with another, possibly dynamically generated, scale 
as well as the fact that the potential is very flat at large values of the bosonic field suggests that the effective 
theory explored here and more generally 
theories emerging from the deformed superspace \cite{Seiberg:2003yz} when coupled to gravity might be of some interest for models of inflation 
\cite{Linde:1990nc}. Again we stress that we are not really entitled to continue the theory in the Minkowski space and our example is just meant to be a primitive attempt in the quest for possible physical applications. There are however already interesting attempts \cite{Chaichian:2003dp} to define superspace deformations directly in the Minkowski space.

The paper is organized as follows: In section two we briefly review the basic properties defining the noncommuting ${\cal N}=1/2$ 
superspace. In section three we investigate the product of a generic number of (anti)chiral superfields in the deformed space. In the 
fourth section we extend the generic supersymmetric chiral sigma model to the deformed superspace while providing a number of useful relations. 
In section five we summarize the result for the generic deformed superfield expressed in terms of the component fields. In this section we also 
confront our Lagrangian with previous results already present in the literature (only for a single field) and show how differences emerge. 
In section six we rewrite the full Lagrangian in a compact form. 

In section seven and using a simple K\"{a}hler we 
compute the potential of the theory after having integrated out the auxiliary fields. The potential is, in general, a series in the deformation 
parameter of the superspace and is, in general, complex. We then present an example and indicate possible physical applications. Section eight is dedicated to the conclusions. Three appendices are dedicated either to relevant examples or lengthy but detailed computations useful for the interested reader 
to better understand our results.

\section{$\mathcal{N}=\frac{1}{2}$ Superspace Preliminaries}
\label{sec:superspace}
Our starting point will be the deformation of ordinary $\mathcal{N}=1$
superspace by  requiring $\theta^{\alpha}$ to
satisfy the following anticommutator relation\footnote{We use the Wess and Bagger notation \cite{WB}.}:
\begin{equation}
  \label{eq:1}
  \{\theta^{\alpha},\theta^{\beta}\}=C^{\alpha\beta} \ .
\end{equation}
The first implications of such an extension were investigated in \cite{Seiberg:2003yz}. 
The deformation requires an ordering definition for the functions of $\theta$. Weyl ordering seems 
to be a natural choice \cite{Seiberg:2003yz}. In this
way every expression containing $\theta$ has to be ordered. This can be
 achieved, following ordinary noncommutative geometry, by introducing the 
star product:
\begin{eqnarray}
  \label{eq:2}
  f(\theta)\ast g(\theta) & = & f(\theta)\exp\bigg(-\frac{C^{\alpha\beta}}{2}
  \overleftarrow{\frac{\partial}{\partial
  {\theta}^{\alpha}}}\overrightarrow{\frac{\partial}{\partial{\theta}^
  {\beta}}} \bigg)g(\theta) \nonumber
  \\
  & = & f(\theta)\bigg(1-\frac{C^{\alpha \beta}}{2}
  \overleftarrow{\frac{\partial}{\partial
  {\theta}^{\alpha}}}\overrightarrow{\frac{\partial}{\partial{\theta}^
  {\beta}}}-\text{det}C \overleftarrow{\frac{\partial}{\partial \theta 
  \theta}}\overrightarrow{\frac{\partial}{\partial \theta
  \theta}}\bigg)g(\theta) \ .
\end{eqnarray}

For the remaining coordinates a simple consistent choice of the (anti)commuting relations is \cite{Seiberg:2003yz}: 

\begin{displaymath}
   \{\theta^{\alpha},\theta^{\beta}\} = C^{\alpha\beta} \ , \nonumber
\end{displaymath}
\begin{equation} 
   \label{eq:3}
   \{{\overline{\theta}}^{\dot{\alpha}},{\overline{\theta}}^{\dot{\beta}}
   \} = \{ {\overline{\theta}}^{\dot{\alpha}},{\theta}^{\beta}
   \} = 0 \ , 
\end{equation}

\begin{displaymath} 
   [y^{\mu},y^{\nu}] = [y^{\mu},{\theta}^{\alpha}] = 
   [y^{\mu},{\overline{\theta}}^{\dot{\alpha}}]=0 \ , 
\end{displaymath}
where $y^{\mu}=x^{\mu}+i\theta\sigma^{\mu}\overline{\theta}$ is the
chiral coordinate. One can imagine various generalizations of this superspace 
which will not be considered here.

Several remarks are in order. Since the
$\overline{\theta}$'s are taken to anticommute it is clear that
they cannot be the complex conjugate of $\theta$. We therefore
have to be in the Euclidean space where the Lorentz group becomes
$SO(4)=SU_{L}(2)\times SU_{R}(2)$. We will however keep the
Lorentz notation. According to \cite{Seiberg:2003yz}
the chiral coordinates $y^{\mu}$ is assumed to commute. Consistency then requires the ordinary
spacetime coordinates $x^{\mu}$ not to commute. In fact, using the relations:

\begin{equation}
  \label{eq:4}
  \theta\sigma^{\mu}\overline{\theta}\ast\theta\sigma^{\nu}\overline{\theta}
  = -\frac{1}{2}\theta\theta\overline{\theta}\overline{\theta}\eta^{\mu\nu}
  -\frac{1}{2}\overline{\theta}\overline{\theta}C^{\mu\nu} \ ,
\end{equation}
\begin{equation}
  \label{eq:5}
  C^{\mu\nu} \equiv C^{\alpha\beta}{\varepsilon}_{\beta\gamma}
  (\sigma^{\mu\nu})_{\alpha}^{\phantom{\alpha}\gamma}=-C^{\nu\mu} \ ,
\end{equation}
we obtain:
\begin{equation}
  \label{eq:6}
  [x^{\mu},x^{\nu}]=\overline{\theta}\overline{\theta}C^{\mu\nu} \ .
\end{equation}
 The reason for \emph{not} assuming the ordinary spacetime coordinates
$x^{\mu}$ to commute is that it makes it hard to define chiral
superfields. Recall that a chiral superfield $\Phi$ is defined through the standard
constraint $\overline{D}_{\dot{\alpha}}\Phi =0$ where
$\overline{D}_{\dot{\alpha}}$ is the covariant
derivative. In the $x^{\mu}$ 
coordinates $\overline{D}_{\dot{\alpha}}$ also depends on $\theta$ and therefore, 
in the deformed superspace, it does not act as an ordinary derivative, i.e. it does not satisfy the
Leibniz rule. In the chiral coordinates $y^{\mu}$ on the contrary,
$\overline{D}_{\dot{\alpha}}=-\frac{\partial}{\partial\overline{\theta}^
{\dot{\alpha}}}$ does not depend explicitly on $\theta$ and therefore acts as an
ordinary derivative. So the deformed superspace is best parameterized by
the following set of coordinates $(y^{\mu},\theta^{\alpha},\overline{\theta}^{\dot
{\alpha}})$ satisfying the algebra (\ref{eq:3}). 

The deformed superspace we refer to as the noncommutative
space. It is clear that for vanishing $C$ this becomes the commutative
space. Any function
defined on the noncommutative space is understood in terms of its
Taylor expansion but with every product replaced by the star
product. The latter breaks half of the
supersymmetry but preserves chirality, i.e. the star product of two
(anti)chiral superfields is again a (anti)chiral superfield. 

It is convenient for the reader to review the expressions for the covariant derivatives $D_{\alpha}$,
$\overline{D}_{\dot{\alpha}}$ and the supercharges
$Q_{\alpha}$, $\overline{Q}_{\dot{\alpha}}$ in the standard way:

\begin{eqnarray}
  D_{\alpha} & = & \frac{\partial}{\partial
  {\theta}^{\alpha}}+2i\sigma^{\mu}_{\alpha \dot{\alpha}}
  {\overline{\theta}}^{\dot{\alpha}}\frac{\partial}{\partial y^{\mu}}
  \ ,
  \label{eq:7} 
  \\ 
  {\overline D}_{\dot{\alpha}} & = & -\frac{\partial}{\partial
  \overline{\theta}^{\dot{\alpha}}} \ , 
  \label{eq:8}
  \\
  Q_{\alpha} & = & \frac{\partial}{\partial {\theta}^{\alpha}} \ ,
  \label{eq:9}
  \\
  {\overline Q}_{\dot{\alpha}} & = & -\frac{\partial}{\partial
  \overline{\theta}^{\dot{\alpha}}}+2i{\theta}^{\alpha}\sigma^{\mu}_
  {\alpha\dot{\alpha}}\frac{\partial}{\partial y^{\mu}} \ .
  \label{eq:10}
\end{eqnarray}

Note that $\overline{Q}_{\dot{\alpha}}$ depends explicitly on $\theta$. 
Hence while all of the anticommutation relations are as in
the commutative space we have for the $\overline{Q}$:

\begin{equation}
  \label{eq:11}
  \big\{\overline{Q}_{\dot{\alpha}},\overline{Q}_{\dot{\beta}}\big\}=-4C^
  {\alpha\beta}\sigma^{\mu}_{\alpha\dot{\alpha}}\sigma^{\nu}_{\beta\dot{\beta}}
  \frac{\partial^{2}}{\partial y^{\mu}\partial y^{\nu}} \ . 
\end{equation}
This shows that the deformed superspace is generated by the supercharge
$Q$ and that the star product is invariant under $Q$. However since $\overline{Q}$ contains a $\theta$ the star
product is not invariant under $\overline{Q}$. So only transformations induced by $Q$ are still 
intact symmetries while the ones induced by $\overline{Q}$ are broken. For this reason the deformed
superspace has been termed $\mathcal{N}=\frac{1}{2}$ superspace. 

\section{(Anti)Chiral Superfields Properties}
\label{sec:chiral-superfields}

\subsection{Chiral Superfields}
Since chirality is respected even in the deformed superspace we define chiral
superfields through the standard condition
$\overline{D}_{\dot{\alpha}}\Phi=0$ which leads to
$\Phi=\Phi(y,\theta)$. Using the fact that
$\theta^{\alpha}\ast\theta_{\alpha}=\theta\theta$ we write:

\begin{equation}
  \label{eq:12}
  \Phi(y,\theta)=A(y)+\sqrt{2}\theta\psi(y)+\theta\theta F(y) \ ,
\end{equation}

The star product of two chiral superfields reads:

\begin{eqnarray}
  \label{eq:13}
  \Phi_1(y,\theta)\ast\Phi_2(y,\theta) & = & \Phi_1(y,\theta)\Phi_2(y,\theta)-
  \frac{C^{\alpha\beta}}{2}\frac{\partial}{\partial\theta^{\alpha}}\Phi_1
  (y,\theta)\frac{\partial}{\partial\theta^{\beta}}\Phi_2(y,\theta)
  \nonumber
  \\
  & & -\det C
  \frac{\partial}{\partial\theta\theta}\Phi_1(y,\theta)\frac{\partial}
  {\partial\theta\theta}\Phi_2(y,\theta) \ .
\end{eqnarray}
By construction the star product respects chirality, i.e. $\Phi_1\ast\Phi_2$ is
again a chiral superfield. This fact allows to define the superpotential as a product of chiral 
superfields. Ambiguities in extending the ${\cal N}=1$ superpotentials to the deformed superspace arise due to the fact that in the noncommutative theory: $\Phi_1 \ast \Phi_2 \neq \Phi_2 \ast \Phi_1$. Different 
noncommutatitive theories correspond to the same commutative theory. This problem is analogous to the 
one arising when trying to quantize a classical theory. Following the procedure used 
when quantizing a classical theory, in \cite{Seiberg:2003yz} a possible generalization 
to the ${\cal N}=1/2$ case of the product of chiral superfields is that 
every product is always fully symmetrized:
\begin{eqnarray}
  \label{eq:14}
  \Phi_1\ast\Phi_2\vert_s & \equiv & \frac{1}{2}(\Phi_1\ast\Phi_2+
  \Phi_2\ast\Phi_1)
  \nonumber 
  \\
  & = & \Phi_1\Phi_2-\det C
  \frac{\partial}{\partial\theta\theta}\Phi_1
  \frac{\partial}{\partial\theta\theta}\Phi_2 \ . 
\end{eqnarray}
The symmetrized product (\ref{eq:14}) also has the nice 
property that the result only depends on the deformation parameter $C$ through $\det
C$. This is due to the fact that $C^{\alpha\beta}=C^{\beta\alpha}$ is
symmetric and $\frac{\partial}{\partial\theta^{\alpha}}$ is Grassmannian in nature.
As we will see, and already pointed out in \cite{Seiberg:2003yz}, this feature survives when considering 
the symmetrized star product of $n$ different
chiral superfields. 
Hence while the deformation per se breaks Lorentz invariance the resulting theories remain
Lorentz covariant \cite{Seiberg:2003yz}.

In the case of $n$ different chiral superfields we 
symmetrize them according to:
\begin{equation}
  \label{eq:15}
  \Phi_1\ast\ldots\ast\Phi_n\vert_s=\frac{1}{n!}\big(\Phi_1\ast\ldots\ast
  \Phi_n+\text{Perm.}\big) \ .
\end{equation}
 
 We explicitly show the first few examples of the symmetrized star product of chiral 
superfields in appendix A. When analyzing the examples one can immediately observe that
a series in $(-\det C)^j$ with $1 \leq j \leq [\frac{n}{2}]$ is generated\footnote{$[\frac{n}{2}]$ denotes the integer part of
$\frac{n}{2}$.}. One also obtains a
polynomial in $\Phi_i$ with $1 \leq i \leq n-2j$. We stress that 
the only dependence on $\theta$ is via this polynomial. Terms containing explicitly
 $C^{\alpha\beta}$, rather than ${\rm det}C$, will never appear. This is so since such a term would be of
 the form:
\begin{equation}
  \label{eq:23}
  C^{\alpha\beta}\frac{\partial}{\partial\theta^{\alpha}}(\Phi_r\cdots\Phi_s)
  \frac{\partial}{\partial\theta^{\beta}}(\Phi_t\cdots\Phi_u) \ ,
\end{equation}
which vanishes upon symmetrization of the chiral superfields.

Using appendix A as a guide it is not hard to guess, and then prove by induction, that the generic product of $n$ chiral superfields is:
\begin{equation}
  \label{eq:16}
  \Phi_1\ast\ldots\ast\Phi_n\vert_s = \Phi_1\cdot\ldots\cdot\Phi_n
  +\sum_{j=1}^{[\frac{n}{2}]}(-\det
  C)^jb_j(n)\prod_{k=n-j+1}^{n}\frac{\partial}{\partial\theta\theta}\Phi_k
  \vert_s \ ,
\end{equation}
where $b_j(n)$ is a polynomial in the fields of the form:
\begin{equation}
  \label{eq:24}
  b_j(n)=\sum_{i=0}^{n-2j}a_i(j,n)\Phi_0\cdots\Phi_i \ ,\qquad\qquad \Phi_0
  \equiv 1
\end{equation}
and the coefficients $a_i(j,n)$ are constructed as follows: Consider $j$
operators $\frac{\partial}{\partial\theta\theta}$. Then the
coefficient $a_i(j,n)$ is the sum of all possible terms of the form:
\begin{equation}
  \label{eq:25}
  \underbrace{\frac{\partial}{\partial\theta\theta}(\Phi_{i+1}\cdots\Phi_{l_1})
  \frac{\partial}{\partial\theta\theta}(\Phi_{l_1+1}\cdots\Phi_{l_2})
 \cdots\frac{\partial}{\partial\theta\theta}
  (\Phi_{l_{j-1}+1}\cdots\Phi_{n-j})}_{j\ \text{F-terms}} \ ,
\end{equation}
where the superfields are ordered according to
$\Phi_{i+1}\Phi_{i+2}\cdots\Phi_{n-j}$. Unfortunately $b_j(n)$ is a rather complicated function to work with. 
However it has been shown in \cite{Azorkina:2005mx} that it is much more convenient to expand a
generic function of superfields around their bosonic component. In this way they were able to construct the ${\cal N}=1/2$ chiral model with a single field. As stated at the beginning we will address the problem of how to generalize the ${\cal N}=1/2$ model to contain an arbitrary number of chiral superfields. This will greatly simplify not only the construction of the superpotential but also the one of the K\"{a}hler.

\subsection{Antichiral Superfields}
\label{sec:antich-superf}

The antichiral superfields are defined via
the standard condition $D_{\alpha}\overline{\Phi}=0$. Therefore 
$\overline{\Phi}=\overline{\Phi}(\overline{y},\overline{\theta})$
where
$\overline{y}^{\mu}=y^{\mu}-2i\theta\sigma^{\mu}\overline{\theta}$.
It turns out that:
\begin{equation}
  \label{eq:19}
  [\overline{y}^{\mu},\overline{y}^{\nu}]=4\overline{\theta}\overline{\theta}
  C^{\mu\nu} \ .
\end{equation}

We therefore have to order the $\overline{y}^{\mu}$'s and multiply
antichiral superfields according to:

\begin{eqnarray}
  \label{eq:20}
  \overline{\Phi}_1(\overline{y},\overline{\theta})\ast\overline{\Phi}_2
  (\overline{y},\overline{\theta}) & = &
  \overline{\Phi}_1(\overline{y},\overline{\theta})\exp\bigg(2\overline{\theta
  \theta}C^{\mu\nu}\overleftarrow{\frac{\partial}{\partial
  {\overline y}^{\mu}}}\overrightarrow{\frac{\partial}{\partial{\overline
  y}^{\nu}}} \bigg) \overline{\Phi}_2(\overline{y},\overline{\theta})
  \nonumber
  \\
  & = & \overline{\Phi}_1(\overline{y},\overline{\theta})\overline{\Phi}_2
  (\overline{y},\overline{\theta})+2\overline{\theta\theta}C^{\mu\nu}
  \frac{\partial}{\partial{\overline y}^{\mu}}\overline{\Phi}_1 (\overline{y},
  \overline{\theta})\frac{\partial}{\partial{\overline y}^{\nu}} 
  \overline{\Phi}_2(\overline{y},\overline{\theta}) \ .\nonumber \\ 
\end{eqnarray}

As in the case of just chiral superfields we will symmetrize with respect to all of the antichiral superfields. Since
$C^{\mu\nu}=-C^{\nu\mu}$ is antisymmetric we obtain:

\begin{eqnarray}
  \label{eq:21}
  \overline{\Phi}_1\ast\overline{\Phi}_2\vert_{\overline{s}} & = &
  \frac{1}{2}(\overline{\Phi}_1\ast\overline{\Phi}_2+\overline{\Phi}_2\ast
  \overline{\Phi}_1)
  \nonumber
  \\
  & = & \overline{\Phi}_1\overline{\Phi}_2 \ , 
\end{eqnarray}
i.e. there are no corrections due to the star product of antichiral
superfields with respect to the ordinary product. This is also true for the star product of $n$ different antichiral
superfields: 

\begin{equation}
  \label{eq:22}
  \overline{\Phi}_1\ast\ldots\ast\overline{\Phi}_n\vert_{\overline{s}}=
  \overline{\Phi}_1\cdot\ldots\cdot\overline{\Phi}_n \ .
\end{equation}
The previous identity can be proved using induction.

The situation becomes more involved when one needs to multiply fields with opposite chirality. In this case 
it is necessary to expand the antichiral superfield as follows:

\begin{eqnarray}
  \label{eq:26}
  \overline{\Phi}(y-2i\theta\sigma\overline{\theta},\overline{\theta})
  & = &
  \overline{A}(y)+\sqrt{2}\overline{\theta}\overline{\psi}(y)+\overline{\theta}
  \overline{\theta}\overline{F}(y)
  \nonumber
  \\
  & &
  +\sqrt{2}\theta\bigg(i\sigma^{\mu}\partial_{\mu}\overline{\psi}(y)
  \overline{\theta}\overline{\theta}-i\sqrt{2}\sigma^{\mu}\overline{\theta}
  \partial_{\mu}\overline{A}(y)\bigg)+\theta\theta\overline{\theta}\overline
  {\theta}\partial^2 \overline{A}(y) \ ,\nonumber \\ 
\end{eqnarray}
 and multiply them according to:

 \begin{eqnarray}
   \label{eq:27}
   \Phi(y,\theta)\ast\overline{\Phi}(y-2i\theta\sigma\overline{\theta},
   \overline{\theta})
   & = & \Phi(y,\theta)\exp\bigg(-\frac{C^{\alpha\beta}}{2}
   \overleftarrow{\frac{\partial}{\partial
   {\theta}^{\alpha}}}\overrightarrow{\frac{\partial}{\partial{\theta}^
   {\beta}}}
   \bigg)\overline{\Phi}(y-2i\theta\sigma\overline{\theta},\overline
   {\theta}) \ . \nonumber \\
 \end{eqnarray}

\section{Extension of the Chiral Sigma Model to ${\cal N}=1/2$.}
\label{sec:sigma-model}

The main goal is to construct the most general chiral sigma model Lagrangian in
$\mathcal{N}=\frac{1}{2}$ superspace featuring
an arbitrary number of different (anti)chiral superfields. The starting point is the generic chiral 
sigma model in four spacetime dimensions in ${\cal N}=1$
which is usually represented as the sum of a K\"{a}hler term and a superpotential:
\begin{equation}
  \label{eq:30}
  \mathcal{L}=\int d^4\theta K(\Phi^i,\overline{\Phi}^j)+\int
  d^2\theta P(\Phi^i)+\int d^2\overline{\theta} \overline{P}(\overline{\Phi}^j) \ .
\end{equation}
We then deform the superspace and indicate this deformation by adding a star label for the superpotential and the K\"{a}hler: 
\begin{equation}
  \label{eq:30}
  \mathcal{L}=\int d^4\theta K_{\ast}(\Phi^i,\overline{\Phi}^j)+\int
  d^2\theta P_{\ast}(\Phi^i)+\int d^2\overline{\theta}
  \overline{P}_{\ast}(\overline{\Phi}^j) \ .
\end{equation}
In the deformed $\mathcal{N}=\frac{1}{2}$ superspace we understand the previous expression in terms 
of a series with every standard product replaced by a star
product. 

\subsection{The Superpotential}

We focus our attention first on the deformed superpotential. Following \cite{Azorkina:2005mx} we choose
to expand around the bosonic component fields
$\{A_i\}$ of the chiral superfields. Therefore we define: 
\begin{eqnarray}
L^i(y,\theta) \equiv \Phi^i(y,\theta)-A^i(y) \ ,
\end{eqnarray} 
 and expand the deformed superpotential according to:
\begin{eqnarray}
  \label{eq:31}
  P_{\ast}(\Phi^i) & = & P(A^i)+L^iP_{,i}+\frac{1}{2!}(L^{i_1}\ast
  L^{i_2})\vert_s
  P_{,i_1i_2}+\ldots
  \nonumber
  \\
  & & +\frac{1}{n!}(L^{i_1}\ast\ldots\ast
  L^{i_n})\vert_sP_{,i_1\ldots i_n}+\ldots \ ,
\end{eqnarray}
where the expansion coefficients are given by:
\begin{equation}
  \label{eq:32}
  P_{,i_1\ldots i_n}\equiv \frac{\partial^n}{\partial
  \Phi^{i_1}\cdots\partial \Phi^{i_n}}P(\Phi)\vert_{\Phi^i = A^i} \ .
\end{equation}

We now recall (\ref{eq:16}) but with every chiral superfield replaced by the associated $L_i$. To fully determine such an expression 
we need to determine the generic product:
\begin{eqnarray}
L^{i_1}\ast\ldots\ast L^{i_n}\vert_s \ .
\end{eqnarray}

More specifically we need to determine $b_j(n)$. Having subtracted the scalar component leads to a large number of simplifications. For example:

\begin{eqnarray} 
  L^{i}L^{j}=-\theta\theta\psi^i\psi^j \ ,
\end{eqnarray} 
and therefore $L^iL^jL^k=0$. This means that
at maximum two L's can be next to each other in $b_j(n)$.
Further cancellations appear 
due to the symmetrization in the fields. Namely the following constraints hold:

\begin{eqnarray}
  \label{eq:33}
  L^iL^j\frac{\partial}{\partial\theta\theta}L^kL^l\vert_s
  & = & \theta\theta(\psi^i\psi^j)(\psi^k\psi^l)\vert_s \nonumber \\
  & \sim &
  (\psi^i\psi^j)(\psi^k\psi^l)+(\psi^i\psi^k)(\psi^j\psi^l)+(\psi^i\psi^l)(\psi^j\psi^k) \nonumber
  \\
 & = & 0 \ , \\
\frac{\partial}{\partial\theta\theta}L^iL^j\frac{\partial}{\partial\theta\theta}L^kL^l\vert_s
  & = & 0 \ . \label{eq:34}
\end{eqnarray}

By definition $b_j(n)$ contains $j$ operators
$\frac{\partial}{\partial\theta\theta}$ and $n-j$ fields
$L^i$. Because of the above constraints the number of fields can be no
larger than the number of operators plus two: $n-j\leq j+2$. So, in
order to obtain non-vanishing contributions we must have:
\begin{equation}
  \label{eq:35}
  \frac{n}{2}-1\leq j \leq \bigg[\frac{n}{2}\bigg] \ .
\end{equation}
If $n$ is even we get non-vanishing contributions for
$j=\frac{n}{2}-1$ and $j=\frac{n}{2}$ and if $n$ is odd we get
non-vanishing contributions only for $j=\frac{n-1}{2}$. If $n$ is even
a bit of algebra yields:

\begin{eqnarray}
  \label{eq:36}
  L^{i_1}\ast\ldots\ast L^{i_{2m}}\vert_s & = & (-\det
  C)^{m-1}\bigg\{L^{i_1}L^{i_2}\frac{\partial}{\partial\theta\theta}L^{i_3}
  \nonumber
  \\
  & & +(m-1)L^{i_1}\frac{\partial}{\partial\theta\theta}(L^{i_2}L^{i_3})\bigg\}
  \prod_{k=4}^{2m}\frac{\partial}{\partial\theta\theta}L^{i_k}\vert_s
  \nonumber \\
  & & + (-\det
  C)^m\prod_{k=1}^{2m}\frac{\partial}{\partial\theta\theta}L^{i_k}
  \nonumber \\
  & = & (-\det
  C)^{m-1}mL^{i_1}L^{i_2}\prod_{k=3}^{2m}\frac{\partial}{\partial\theta\theta}
  L^{i_k}\vert_s
  \nonumber \\ 
  & & +(-\det
  C)^m\prod_{k=1}^{2m}\frac{\partial}{\partial\theta\theta}L^{i_k} 
  \nonumber\\
   & = & -\theta\theta(-\det
  C)^{m-1}\frac{m}{(2m)!}\big\{\psi^{i_1}\psi^{i_2}
  F^{i_3}\cdots F^{i_{2m}}+\text{Perm.}\big\} 
  \nonumber \\
  & &+(-\det C)^mF^{i_1}\cdots 
  F^{i_{2m}} \ ,\nonumber\\
\end{eqnarray}
and if $n$ is odd: 
\begin{eqnarray}
  \label{eq:41}
  L^{i_1}\ast\ldots\ast L^{i_{2m+1}}\vert_s & = & 
  (-\det
  C)^m\bigg\{L^{i_1}\frac{\partial}{\partial\theta\theta}L^{i_2}
  \nonumber \\
  & & +m\frac
  {\partial}{\partial\theta\theta}(L^{i_1}L^{i_2})\bigg\}\prod_{k=3}^{2m+1}
  \frac{\partial}{\partial\theta\theta}L^{i_k}\vert_s \nonumber\\
  & = & (-\det C)^m\bigg\{\theta\theta F^{i_1}\cdots F^{i_{2m+1}}
  \nonumber \\  
  & & +\frac{\sqrt{2}}{(2m+1)!}\theta\big(\psi^{i_1}F^{i_2}\cdots
  F^{i_{2m+1}}+\text{Perm.}\big) \nonumber \\
  & & -\frac{m}{(2m+1)!}\big(\psi^{i_1}\psi^{i_2}F^{i_3}\cdots
  F^{i_{2m+1}} +\text{Perm.}\big)\bigg\} \ . \nonumber \\
\end{eqnarray}
The most general superpotential in $\mathcal{N}=\frac{1}{2}$ superspace is then:
\begin{eqnarray}
  \label{eq:37}
  \mathcal{L}_{pot} & = & \int d^2\theta P(\Phi^i)+\int d^2\overline{\theta}
  \overline{P}(\overline{\Phi}^j) \nonumber \\
  & & +\sum_{n=1}^{\infty}\frac{1}{(2n+1)!}(-\det C)^nF^{i_1}\cdots 
  F^{i_{2n+1}}P_{,i_1\cdots i_{2n+1}} \nonumber \\
  & & -\sum_{n=2}^{\infty}\frac{n}{(2n)!}(-\det
  C)^{n-1}\psi^{i_1} \psi^{i_2}F^{i_3}\cdots
  F^{i_{2n}}P_{,i_1\cdots i_{2n}} \ ,
\end{eqnarray}
where we have used the fact that $P_{,i_1\cdots i_{2n}}$ is symmetric
in all its indices. 

\subsection{The K\"ahler potential }
The situation is considerably more involved for the explicit determination of the K\"ahler potential. 
We now have also to consider antichiral superfields and their product with chiral superfields. We start by
defining 
\begin{eqnarray}\overline{L}^j(y-2i\theta\sigma\overline{\theta},\overline{\theta}) 
\equiv \overline{\Phi}^j(y-2i\theta\sigma\overline{\theta},\overline{\theta})
-\overline{A}^j(y) \ ,\end{eqnarray} which leads to the immediate constraint:
\begin{eqnarray}\overline{L}^{j_k}\ast\overline{L}^{j_l}\ast\overline{L}^{j_m}=0 \ .\end{eqnarray} In components
\begin{equation}
  \label{eq:40}
  \overline{L}= \sqrt{2}\overline{\theta}\overline{\psi}(y) +
  \overline{\theta}\overline{\theta} 
  \overline{F}(y)+\sqrt{2}\theta\big(i\sigma^{\mu}\partial_{\mu}\overline{\psi}
  (y)\overline{\theta}\overline{\theta}-i\sqrt{2}\sigma^{\mu}\overline{\theta
  }\partial_{\mu}\overline{A}(y)\big)+\theta\theta\overline{\theta}\overline
  {\theta}\partial^2 \overline{A}(y) \ . 
\end{equation}

The K\"ahler potential admits the following series expansion: 
\begin{eqnarray}
  \label{eq:38}
  K_{\ast}(\Phi^i,\overline{\Phi}^j) & = & K(A^i,\overline{A}^j) + 
  \sum_{n=1}^{\infty}\frac{1}{n!}\big( L^{i_1}\ast\ldots \ast L^{i_n}\big)
  \vert_s K_{,i_1\cdots i_n} 
  \nonumber
  \\
  & & + \overline{L}^j
  K_{,j}+\frac{1}{2}
  \big(\overline{L}^{j_1}\ast\overline{L}^{j_2}\big)\vert_{\overline{s}}
  K_{,j_1 j_2}
  \nonumber
  \\
  & & + \sum_{n=1}^{\infty}\sum_{m=1}^{2}\frac{1}{n!m!}\big(
  L^{i_1}\ast\ldots\ast L^{i_n}
  \nonumber \\
  & & \ast \overline{L}^{j_1}\ast\ldots\ast 
  \overline{L}^{j_m}\big)\vert_{s\overline{s}} K_{,i_1\cdots i_n j_1
  \cdots j_m} \ , \nonumber \\
\end{eqnarray}
where each term is fully symmetrized in the chiral and antichiral superfields. The expansion coefficients are:
\begin{equation}
  \label{eq:39}
  K_{,i_1\cdots i_n j_1
  \cdots j_m} \equiv
  \frac{\partial^{(n+m)}}{\partial\Phi^{i_1}
  \cdots\partial\Phi^{i_n}{\partial\overline{\Phi}}^{j_1}\cdots
  {\partial\overline{\Phi}}^{j_m}}K(\Phi^i,\overline{\Phi}^{j})
  \vert_{\Phi^i = A^i(y),\overline{\Phi}^j = \overline{A}^j(y)} \ .
\end{equation}
The first four terms in (\ref{eq:38}) either vanish or do not give any 
contribution in the deformation parameter $\det C$ to the Lagrangian. Only the last term induces 
corrections due to the deformed superspace. 

We start by analyzing in detail the case $m=1$ in the last term of the series. We will use the trace identity:
\begin{equation}
  \label{eq:28}
  \int d^4\theta (\Phi_1\ast\overline{L}\ast\Phi_2) = 
  \int d^4\theta (\Phi_2\ast\Phi_1\ast \overline{L}) = 
  \int d^4\theta (\Phi_2 \ast\Phi_1)\overline{L} \ ,
\end{equation}
where $\Phi_1$ and $\Phi_2$ are two generic chiral superfields. 
Symmetrizing in the antichiral field $\overline{L}^{j_1}$ and using the
above identity, we obtain:
\begin{equation}
  \label{eq:17}
  \int d^4\theta \big(L^{i_1}\ast\ldots\ast L^{i_n}\ast \overline{L}^{j_1}\big)
  \vert_{s\overline{s}} = 
  \int d^4\theta \big(L^{i_1}\ast\ldots\ast L^{i_n}\big)\vert_s 
  \overline{L}^{j_1} \ .
\end{equation}

{}For $m=1$ we can immediately find using (\ref{eq:17}) together with 
(\ref{eq:36}) and (\ref{eq:41}):
\begin{displaymath}
  \int d^4\theta
  \sum_{n=1}^{\infty}\frac{1}{n!}\big(L^{i_1}\ast\ldots
  \ast L^{i_n}\ast \overline{L}^{j_1}\big)\vert_{s\overline{s}}
  K_{,i_1\cdots i_n j_1}
\end{displaymath}
\begin{displaymath}
  = \int
  d^4\theta \bigg\{ L^{i_1}\overline{L}^{j_1}
  K_{,i_1 j_1}+\frac{1}{2}L^{i_1}L^{i_2}\overline{L}^{j_1}K_{
  ,i_1i_2 j_1}\bigg\}
\end{displaymath}
\begin{displaymath}
  -\sum_{n=2}^{\infty}\frac{n(-\det C)^{n-1}}{(2n)!}\psi^{i_1}\psi^{i_2}
  F^{i_3}\cdots F^{i_{2n}}\overline{F}^{j_1}K_{,i_1\cdots i_{2n} j_1}
\end{displaymath}
\begin{equation}
  \label{eq:18}
  +\sum_{n=1}^{\infty}\frac{(-\det C)^n}{(2n)!}F^{i_1}\cdots
  F^{i_{2n}}\partial^2\overline{A}^{j_1}K_{,i_1\cdots i_{2n}j_1}
\end{equation}
\begin{displaymath}
  +\sum_{n=1}^{\infty}\frac{(-\det C)^n}{(2n+1)!}\bigg(
  F^{i_1}F^{i_2}\overline{F}^{j_1}+
  i\partial_{\mu}\overline{\psi}^{j_1}\overline{\sigma}^{\mu}
  \psi^{i_1}F^{i_2}
\end{displaymath}
\begin{displaymath}
   -n\psi^{i_1}\psi^{i_2}\partial^2 \overline{A}^{j_1}
  \bigg)F^{i_3}\cdots F^{i_{2n+1}}K_{,i_1\cdots
  i_{2n+1} j_1} \ .
\end{displaymath}

We have also used the fact that the expansion coefficients are
symmetric in their unbarred indices. 

What we now have left to address is the case $m=2$. We will proceed in a manner similar to
the $m=1$ case. By first symmetrizing $\overline{L}^{j_2}$ and using
the trace identity we find:
\begin{equation}
  \label{eq:29}
  \int d^4\theta\big(L^{i_1}\ast \ldots \ast L^{i_n}\ast \overline{L}^{j_1}\ast
  \overline{L}^{j_2}\big)\vert_{s\overline{s}} = \int d^4 \theta \big(
  L^{i_1}\ast\ldots\ast L^{i_n}\ast\overline{L}^{j_1}\big)
  \vert_{s\overline{s}} \overline{L}^{j_2} \ .
\end{equation}
The interested reader can find the complete computation, much more involved than for the $m=1$ case in the appendix B. Here
it is sufficient to report the final contribution of this term to the Lagrangian:

\begin{displaymath}
  \int d^4 \theta \sum_{n=1}^{\infty}\frac{1}{n!2}\big(L^{i_1}\ast\cdots 
  \ast L^{i_n}\ast
  \overline{L}^{j_1}\ast\overline{L}^{j_2}\big)\vert_{s\overline
  {s}} K_{,i_1\cdots i_{n} j_1 j_2} 
\end{displaymath}
\begin{displaymath}
  =\int d^4 \theta \bigg\{\frac{1}{2}L^{i_1}\overline{L}^{j_1}
\overline{L}^{j_2}K_{,i_1 j_1 j_2} 
+\frac{1}{4}L^{i_1}L^{i_2}
  \overline{L}^{j_1}
  \overline{L}^{j_2}K_{,i_1 i_2 j_1 j_2}\bigg\} 
\end{displaymath}
\begin{displaymath}
  +\sum_{n=2}^{\infty}\frac{(-\det
  C)^{n-1}}{(2n)!}\frac{n}{2}\overline{\psi}^{j_1}\overline{\psi}^{j_2}\psi^{i_1}
  \psi^{i_2}F^{i_3}\cdots F^{i_{2n}}K_{,i_1\cdots i_{2n} j_1 j_2}
\end{displaymath}
\begin{displaymath}
  +\sum_{n=1}^{\infty}\frac{(-\det
  C)^n}{(2n+1)!}\partial^{\mu}\overline{A}^{j_1}
  \partial_{\mu}\overline{A}^{j_2}
  F^{i_1}\cdots F^{i_{2n}} K_{,i_1\cdots i_{2n} j_1 j_2} 
\end{displaymath}
\begin{equation}
  \label{eq:54}
  -\sum_{n=1}^{\infty}\frac{(-\det
   C)^n}{(2n+1)!}\bigg\{i\psi^{i_1}\sigma^{\mu}
  \overline{\psi}^{j_1}\partial_{\mu}\overline{A}^{j_2}+\frac{1}{2}\overline{\psi}^{j_1}
  \overline{\psi}^{j_2}F^{i_1}\bigg\}F^{i_2}\cdots F^{i_{2n+1}}
  K_{,i_1\cdots i_{2n+1} j_1 j_2} \ .
\end{equation}

 \newpage

 \section{The Full Component Chiral Model Lagrangian}
 We can now summarize in components the generic chiral sigma model Lagrangian for an arbitrary number of chiral superfields. 
 This is achieved by combining (\ref{eq:37}) and (\ref{eq:18}),(\ref{eq:54}) 
as well as (\ref{eq:38}). The deformed theory is then:

\begin{eqnarray}
  \label{eq:55}
  \mathcal{L} & = & \int d^4 \theta K(\Phi^i,\overline{\Phi}^j) + \int
  d^2 \theta P(\Phi^i) + \int
  d^2\overline{\theta}\overline{P}(\overline{\Phi}^j)
  \nonumber
  \\
  & & 
  -\sum_{n=2}^{\infty}\frac{(-\det C)^{n-1}}{(2n)!}n\psi^{i_1}\psi
  ^{i_2}F^{i_3}\cdots F^{i_{2n}}\overline{F}^{j_1} K_{,i_1\cdots
  i_{2n} j_1} 
  \nonumber
  \\
  & & +\sum_{n=1}^{\infty} \frac{(-\det C)^n}{(2n)!}F^{i_1}\cdots
  F^{i_{2n}} \partial^2 \overline{A}^{j_1} K_{,i_1\cdots i_{2n} j_1}
  \nonumber
  \\
  & & +\sum_{n=1}^{\infty}\frac{(-\det C)^n}{(2n+1)!}F^{i_1}\cdots
  F^{i_{2n+1}} \overline{F}^{j_1} K_{,i_1\cdots i_{2n+1} j_1}
  \nonumber
  \\
  & & +\sum_{n=1}^{\infty}\frac{(-\det
  C)^n}{(2n+1)!}i\partial_{\mu}\overline{\psi}^{j_1}\overline{\sigma}^{\mu}
  \psi^{i_1} F^{i_2}\cdots F^{i_{2n+1}} K_{,i_1\cdots i_{2n+1} j_1}
  \nonumber
  \\
  & & -\sum_{n=1}^{\infty}\frac{(-\det
  C)^n}{(2n+1)!}n\partial^2 \overline{A}^{j_1}\psi^{i_1}\psi^{i_2}
  F^{i_3}\cdots F^{i_{2n+1}} K_{,i_1\cdots i_{2n+1} j_1}
  \nonumber
  \\
  & & +\sum_{n=2}^{\infty}\frac{(-\det C)^{n-1}}{(2n)!}\frac{n}{2}
  \overline{\psi}^{j_1} \overline{\psi}^{j_2}{\psi}^{i_1} {\psi}^{i_2}
  F^{i_3}\cdots F^{i_{2n}} K_{,i_1\cdots i_{2n} j_1 j_2} 
  \nonumber
  \\
  & & +\sum_{n=1}^{\infty}\frac{(-\det C)^{n}}{(2n+1)!}
  \partial^{\mu}\overline{A}^{j_1}
  \partial_{\mu}\overline{A}^{j_2}
  F^{i_1}\cdots F^{i_{2n}} K_{,i_1\cdots i_{2n} j_1 j_2} 
  \nonumber
  \\
  & & -\sum_{n=1}^{\infty}\frac{(-\det
  C)^n}{(2n+1)!}i\psi^{i_1}\sigma^{\mu}
  \overline{\psi}^{j_1}\partial_{\mu}\overline{A}^{j_2}F^{i_2}\cdots 
  F^{i_{2n+1}}
  K_{,i_1\cdots i_{2n+1} j_1 j_2}
  \nonumber
  \\
  & & -\sum_{n=1}^{\infty}\frac{(-\det
   C)^n}{(2n+1)!}\frac{1}{2}\overline{\psi}^{j_1}
  \overline{\psi}^{j_2}F^{i_1}\cdots F^{i_{2n+1}}
  K_{,i_1\cdots i_{2n+1} j_1 j_2}
  \nonumber
  \\
  & & +\sum_{n=1}^{\infty}\frac{(-\det C)^n}{(2n+1)!}F^{i_1}\cdots 
  F^{i_{2n+1}}P_{,i_1\cdots i_{2n+1}} 
  \nonumber 
  \\
  & & -\sum_{n=2}^{\infty}\frac{(-\det C)^{n-1}}{(2n)!}n\psi^{i_1} 
  \psi^{i_2}F^{i_3}\cdots F^{i_{2n}}P_{,i_1\cdots i_{2n}} \ .
\end{eqnarray}
This is the complete expression for the chiral model with a generic number of chiral superfields. 
A similar action in the two dimensional deformed superspace framework has been developed by Chandrasekhar and Kumar in \cite{Chandrasekhar:2003uq, Chandrasekhar:2004ti}. As already noticed by Seiberg the theory splits into the $C=0$ Lagrangian and a part depending on 
the deformation parameter via its determinant. The result is quite involved and generalizes the single field result for the four dimensional superspace due to \cite{Azorkina:2005mx,{Hatanaka:2005rg}}. {}A few observations are in order. It is rather hard to integrate out the auxiliary fields without invoking an expansion in the deformation parameter ${\rm det} C$ when the K\"{a}hler is not just a simple quadratic term in the superfields.   

Another subtle point is in the definition of the deformed K\"{a}hler.  When deriving the
previous Lagrangian we have {\it fully} symmetrized the chiral and antichiral superfields in performing the star product
associated to the
expansion of the K\"ahler potential. This seems to lead to few differences with previous results obtained in the literature both for the deformed four dimensional superspace \cite{Azorkina:2005mx} as well as the 
two dimensional one \cite{Chandrasekhar:2003uq},\cite{Chandrasekhar:2004ti}.
The differences might arise due to apparently different definitions 
for the modified K\"{a}hler potential.  
We have already explained that we have symmetrized with respect to both chiral and antichiral fields treating the 
K\"{a}hler. This is the same procedure we have used for the superpotential with different chiral superfields. However one could also imagine 
to symmetrize independently the (anti)chiral superfields per se and then multiply them together just before integrating in the superspace variables. The two procedures lead to different terms emerging in the Lagrangian. Note that to 
observe such differences one has to consider at least two antichiral superfields and more than one chiral field in the K\"{a}hler. 

If we consider the second symmetrization the terms in the Lagrangian modify as follows: The fifth term counting from the bottom becomes:
\begin{eqnarray}
\sum_{n=1}^{\infty}\frac{(-\det C)^{n}}{(2n)!}
  \partial^{\mu}\overline{A}^{j_1}
  \partial_{\mu}\overline{A}^{j_2}
  F^{i_1}\cdots F^{i_{2n}} K_{,i_1\cdots i_{2n} j_1 j_2} \ .
  \end{eqnarray} 
A new term appears which is:
\begin{eqnarray}
-\sum_{n=1}^{\infty}\frac{(-\det C)^{n}}{(2n+1)!}n
  \partial^{\mu}\overline{A}^{j_1}
  \partial_{\mu}\overline{A}^{j_2} \psi^{i_1} \psi^{i_2}
  F^{i_3}\cdots F^{i_{2n+1}} K_{,i_1\cdots i_{2n+1} j_1 j_2} \ .
  \end{eqnarray} 
This would then seem to agree with the results found in \cite{Azorkina:2005mx} when specializing to the single field theory. The limit of a single field can be found in appendix C. 

\section{Expressing the $N=1/2$ Generic Chiral Model in a compact form}
 
Recently it was shown that both, in the
two dimensional case and in the four dimensional case with only one
chiral superfield it is possible to express the above Lagrangian \cite{{Alvarez-Gaume:2005pj},{Azorkina:2005mx}} in 
a more compact from. This is possible in our case as well with the introduction of the following functions:
\begin{equation}
  \label{eq:2v}
  \mathcal{P}(A^i,F^i)=\frac{1}{2}\int_{-1}^{1}d\tau P(A^i+\tau\sqrt{-
  \det C}F^i)
\end{equation}

\begin{equation}
  \label{eq:3v}
  \mathcal{K}(A^i,F^i,\overline{A}^j)=\frac{1}{2}\int_{-1}^{1}d\tau
  K(A^i+\tau\sqrt{- \det C}F^i,\overline{A}^j)
\end{equation}

\begin{equation}
  \label{eq:4v}
  \mathcal{K}'(A^i,F^i,\overline{A}^j)=\frac{1}{2}\int_{-1}^{1}d\tau \tau
  K(A^i+\tau\sqrt{- \det C}F^i,\overline{A}^j)
\end{equation}

\begin{equation}
  \label{eq:5v}
  \mathcal{K}''(A^i,F^i,\overline{A}^j)=\frac{1}{2}\int_{-1}^{1}d\tau
  \frac{\partial}{\partial\tau}\big(\tau
  K(A^i+\tau\sqrt{- \det C}F^i,\overline{A}^j)\big)
\end{equation}
The Lagrangian presented in (\ref{eq:55}) can than be rewritten as:
\begin{eqnarray}
  \label{eq:8v}
  \mathcal{L} & = & \int d^4 \theta
  \mathcal{K}(\Phi^i,F^i,\overline{\Phi}^j) + \int d^2 \theta
  \mathcal{P}(\Phi^i,F^i) + \int
  d^2\overline{\theta}\overline{\mathcal{P}}(\overline{\Phi}^j)
  \nonumber
  \\
  & & -\partial^2\overline{A}^{j_1}\mathcal{K}_{,j_1}  
  -\frac{\sqrt{- \det C}}{2} \partial^2 \overline{A}^{j_1}\psi^{i_1}\psi^{i_2}
  \mathcal{K}'_{,i_1 i_2 j_1} 
  + \partial^2
  \overline{A}^{j_1} \mathcal{K}''_{,j_1}
\end{eqnarray}
Interestingly even though the full Lagrangian can be expressed in a simple
manner it still does not exhibit the standard structure of the original supersymmetric 
chiral model. Specifically there are three terms which - for a generic K\"ahler
potential -  break the K\"ahler
structure. This is perhaps not too surprising since the
$\mathcal{N}=1/2$ deformed superspace is not expected to be a K\"ahler
manifold. 

It might also be instructive to rewrite the previous Lagrangian in a new form in which 
only two terms breaking the K\"ahler structure of the theory appear:\footnote{We have used the identities: $\sqrt{-\det
    C}\frac{\partial}{\partial
    A^i}\mathcal{K}'=\frac{\partial}{\partial F^i}\mathcal{K}$ and
  $\mathcal{K}''=\mathcal{K}+F^i\frac{\partial}{\partial F^i}\mathcal{K}$}

\begin{eqnarray}
  \label{eq:14v}
   \mathcal{L} & = & \int d^4 \theta
  \mathcal{K}(\Phi^i,F^i,\overline{\Phi}^j) + \int d^2 \theta
  \mathcal{P}(\Phi^i,F^i) + \int
  d^2\overline{\theta}\overline{\mathcal{P}}(\overline{\Phi}^j)
  \nonumber
  \\
  & & +\partial^2\overline{A}^{j_1} \bigg( F^{i_1}\frac{\partial}{\partial F^{i_1}}
  \mathcal{K}_{,j_1}-\frac{1}{2}\psi^{i_1}\psi^{i_2}\frac{\partial}{\partial
  F^{i_1}} \mathcal{K}_{,i_2j_1} \bigg) \ .
\end{eqnarray}

As we have already discussed different symmetrization procedures of the fields within the 
K\"ahler potential lead to different terms appearing in the
Lagrangian. In deriving the Lagrangian (\ref{eq:14v}) we fully symmetrized both the
chiral and antichiral superfields. If we choose, however, to symmetrize the
chiral and antichiral superfields independently and then we multiply
them together we obtain, in addition to the Lagrangian (\ref{eq:14v}),
two new terms which spoil the K\"ahler structure of the original supersymmetric theory.  The full Lagrangian assuming this second symmetrization reads: 
\begin{eqnarray}
  \label{eq:15v}
   \mathcal{L} & = & \int d^4 \theta
  \mathcal{K}(\Phi^i,F^i,\overline{\Phi}^j) + \int d^2 \theta
  \mathcal{P}(\Phi^i,F^i) + \int
  d^2\overline{\theta}\overline{P}(\overline{\Phi}^j)
  \nonumber
  \\
  & & +\partial^2\overline{A}^{j_1} \bigg( F^{i_1}\frac{\partial}{\partial F^{i_1}}
  \mathcal{K}_{,j_1}-\frac{1}{2}\psi^{i_1}\psi^{i_2}\frac{\partial}{\partial
  F^{i_1}} \mathcal{K}_{,i_2j_1} \bigg)
  \nonumber
  \\
  & & +
  \partial^{\mu}\overline{A}^{j_1}\partial_{\mu}\overline{A}^{j_2}\bigg(
  F^{i_1} \frac{\partial}{\partial F^{i_1}} \mathcal{K}_{,j_1j_2} -
  \frac{1}{2} \psi^{i_1}\psi^{i_2} \frac{\partial}{\partial F^{i_1}}
  \mathcal{K}_{,i_2j_1j_2} \bigg)
\end{eqnarray}
It is important to stress that this point seems to have been overlooked in the literature.

\section{The Scalar Potential and a Toy Model}
It is hard to solve for the potential of the theory with a general K\"{a}hler. However integrating out 
the auxiliary fields is straightforward
in the case of the standard K\"{a}hler (i.e. made only out of the sum of quadratic terms for each field). In 
this case the potential of the theory can be written explicitly in terms of the scalar components of 
the chiral and antichiral superfields:
\begin{eqnarray}
V= \frac{\partial P}{\partial A^i} \frac{\partial \overline{P}}{\partial \overline{A}^i} 
+\sum_{n=1}^{\infty}\frac{(-{\rm det}C)^n}{(2n+1)!}\frac{\partial \overline{P}}{\partial \overline{A}^{i_1}} \cdots 
\frac{\partial \overline{P}}{\partial \overline{A}^{i_{2n+1}}} P_{,i_1 \cdots i_{2n+1}} \ .
\end{eqnarray}
We recall that we are in the Euclidean space. It is not clear how to continue in the Minkowski space. Interesting attempts to define deformed superspaces 
suitable for Minkowski space are provided in \cite{Chaichian:2003dp}. One could imagine to perform the continuation after the full superspace integration has been performed. What would be the ground state of the theory? The potential is, in general, complex
 and hence it would describe an unstable theory.

We now propose a toy model for which the potential of the theory at the component level, and after having integrated out the auxiliary field, 
contains an infinite number of terms. Interestingly the full series in the deformation parameter can be summed. 
Consider the superpotential:
\begin{eqnarray}
P(\Phi)=\Lambda^3 \log \left(\frac{\Phi}{\Lambda}\right) \ .
\end{eqnarray}  
This superpotential is not renormalizable and should be considered as an effective theory of some more fundamental theory which dynamically generates the energy scale $\Lambda$. 
Summarizing, the ${\cal N}=1$ theory we consider is then:
\begin{eqnarray}
\int d^4\theta\, \overline{\Phi} \Phi + \int d^2\theta \, P(\Phi) +  \int d^2\overline{\theta} \,
\overline{P}(\overline{\Phi}) \ .
\end{eqnarray}
The potential of the theory obtained after replacing the ordinary product with a star product can now be computed and reads:
\begin{eqnarray}
V[A,\bar{A}]= \frac{(\Lambda \bar{\Lambda})^{\frac{3}{2}}}{\sqrt{{\rm det} C}} \arctan \left[\frac{(\Lambda \bar{\Lambda})^{\frac{3}{2}} \sqrt{{\rm det} C}}{
\bar{A}A}\right] \ .
\end{eqnarray}  
An amusing property of the 
final potential is that it is a function only of the product of the scalar and the antiscalar field. Although we are forced, by the very 
definition of the deformed superspace, to be in the Euclidean space the structure of the potential is so intriguing that we have been 
tempted to analytically continue our example in the Minkowski space. 
If we choose ${\rm det }C$ to be real this potential is real and also
positive definite. Regardless of whether ${\det C}$ is positive or negative. In the limit of null deformation 
one recovers the ${\cal N}=1$ supersymmetric potential which has a runaway vacuum at large values of the scalar field $A$.

The main problem with reality is that even if we assume the determinant of the deformation 
parameter $C$ to be real the potential contains terms of the type:
\begin{eqnarray}
\left(\frac{\partial \overline{P}}{\partial \overline{A}}\right)^{2n+1} \frac{\partial^{2n+1} P}{(\partial A)^{2n+1} }\ , 
\end{eqnarray}
where, for simplicity, we are considering the single field theory. These terms are, for a generic superpotential, non 
symmetric under the exchange of $A$ with $\bar{A}$.  
In our example we have shown that it is possible to construct a superpotential which leads to terms in the deformed extension 
of the potential symmetric in the exchange of $A$ with $\bar{A}$. This is a quite distinctive property of the example presented. 
This exchange symmetry becomes the self conjugation property of the potential once continued the model in the Minkowski space.
 
\begin{figure}[htbp] 
\begin{center} 
\includegraphics[scale=.5]{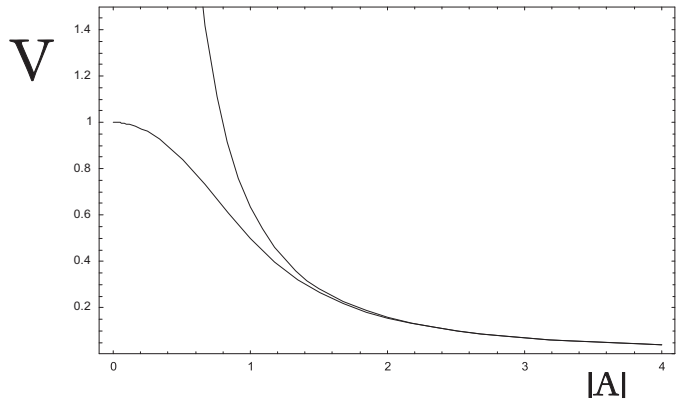} 
\end{center} 
\caption{Potential of the theory normalized to $\pi/2$. The upper curve corresponds to the undeformed ${\cal N}=1$ potential. The lower curve is the potential in the deformed theory with the choice of 
${\rm det} C=1$ in units of $\Lambda$. } 
\label{figure1} 
\end{figure}

It is interesting to compare the behavior of the new potential with the one in the ${\cal N}=1$ theory. At large values of the fields the deformed and undeformed theory share the same properties, i.e. 
the vacuum of the theory is still of a runaway type. The main differences arise near the origin 
of the field space\footnote{{}For completeness we also discuss the case in which $\det C$ is negative. In 
this case the
potential is:
\begin{eqnarray}
V[A,\bar{A}]= \frac{(\Lambda \bar{\Lambda})^{\frac{3}{2}}}{\sqrt{-{\rm det} C}}\tanh^{-1}\left[\frac{(\Lambda \bar{\Lambda})^{\frac{3}{2}} \sqrt{-{\rm det} C}}{
\bar{A}A}\right] \ .
\end{eqnarray}  
The potential is not defined in the following region of field space:
\begin{eqnarray}
\bar{A}A< (\Lambda \bar{\Lambda})^{\frac{3}{2}} \sqrt{-{\rm det} C} \ ,
\end{eqnarray}
which must then be excluded. This seems a rather unphysical situation.}. Indeed for any finite and positive ${\rm det} C$ the potential 
develops a maximum at the origin of the field space with the finite value:
\begin{eqnarray}
V[A=0] = \frac{\pi}{2}\frac{|\Lambda|^3}{\sqrt{{\rm det} C}} \ . 
\end{eqnarray}  
This must be contrasted with the original undeformed potential which diverges at the origin of the $A$ field. We present the potential of the theory in fig.~\ref{figure1}. 
Another property of this potential is the natural presence of two scales, i.e. the dynamically generated one $\Lambda$ due perhaps to some strongly interacting underlying gauge theory\footnote{Note 
that the present effective theory resembles, although it is not the same, the Veneziano Yankielowicz (VY) Theory \cite{Veneziano:1982ah} which has already been investigated in the deformed superspace framework in \cite{Hatanaka:2005rg} with a simple K\"{a}hler. It would be interesting also to investigate in the deformed superspace the extended VY theory \cite{Merlatti:2004df}. 
The extension is needed to be able to account for a number of relevant properties linked 
to the vacuum structure of super Yang-Mills \cite{Merlatti:2004df}.} and the one associated to the deformation parameter $C$ which can be
linked (see \cite{Seiberg:2003yz}) to a nontrivial graviphoton background.  

The noncommutative space modifies, at the potential level, the low energy behavior of the theory. It naturally introduces a new maximum with a finite value of the potential while the theory still preserves Lorentz symmetry \cite{Seiberg:2003yz}. 

The presence of the gravity scale together with another dynamically generated scale 
as well as the fact that the potential is very flat suggests that the effective theory explored here and more generally 
theories emerging from the deformed superspace \cite{Seiberg:2003yz}, after coupling with gravity, could potentially lead to interesting models of inflation when trying 
to identify the inflaton with the field $A$. Interestingly our toy model provides an example of the so called natural type \cite{Linde:1990nc} of inflationary models which, however, does not emerge due 
to spontaneous symmetry breaking but due to an underlying superspace deformation. 

\section{Conclusions}
We have constructed the component 
form of the most general chiral model theory with an arbitrary number of chiral superfields in 
the deformed ${\cal N} = 1/2$ superspace theory in four dimensions. This generalizes the work in \cite{Azorkina:2005mx} presented for a single chiral superfield. Due to the fact that the product of chiral superfields 
does not commute in the deformed superspace we used, following \cite{Seiberg:2003yz}, an ordering procedure which renders the 
generalization to an arbitrary number of fields technically involved. 

We have also elucidated some issues related to how the K\"{a}hler of the theory is handled. More precisely we show that certain terms, present in \cite{Azorkina:2005mx}, are actually absent when considering the fully symmetrized deformed K\"{a}hler of the theory with respect to both chiral and antichiral superfields simultaneously. 

Having determined the general chiral model for $N=1/2$ with an arbitrary number of superfields will allow 
to investigate the deformations of a number of chiral models derived in the ${\cal N}=1$ supersymmetric contest. It would also be interesting, in the future, to investigate the phase structure of 
these theories \cite{Gubser:2000cd,Ambjorn:2002nj}.

We then proposed a toy model in which the potential of the theory at the component level
contains an infinite number of terms. Interestingly the full series in the deformation parameter can be summed. Another amusing property of the 
final potential is that it is a function only of the product of the scalar and the antiscalar field. Although we are forced, by the present
definition of the deformed superspace, to be in the Euclidean space the structure of the potential is so intriguing to us that we analytically continued our example in the Minkowski space. Here we have shown that, if the determinant of the deformation parameter is real and positive, 
also the potential is real and positive. The vacuum structure of the theory is unchanged with respect to the undeformed space while the 
potential is largely affected in the infrared, i.e. for small values of the bosonic field values. The deformation naturally introduces a new maximum with a finite value of the potential at the origin of the field space while the theory still preserves Lorentz symmetry \cite{Seiberg:2003yz}. 

The presence of two scales, i.e. the one associated to the deformation parameter, naturally linked to the the gravity scale, 
together with another (possibly dynamically generated) scale 
as well as the fact that the potential is very flat at large values of the bosonic field suggests 
that the toy model explored here and more generally 
theories emerging from the deformed superspace \cite{Seiberg:2003yz} may lead to interesting models of inflation. 

Our results may also be useful when investigating central extensions and/or domain wall solutions \cite{Chu:2005me} for general chiral models in 
$N=1/2$.

\subsubsection*{Note Added In Proof}
After our paper was submitted to the archive a related paper \cite{Hatanaka:2005fx} appeared. In our original 
submission we did not present the compact form of the Lagrangian.  

\vskip .5cm \centerline{\bf Acknowledgments} 
\vskip .5cm
We thank I.L. Buchbinder, B. Chandrasekhar, P. Di Vecchia, D.D. Dietrich, L. Griguolo, J. Grimstrup, A. Jokinen, P. Merlatti, J. Schechter, M. Shifman and G. Vallone for careful reading of the manuscript or discussions.  
 
 \newpage
\appendix

\section{Examples of Symmetrized Star Products} \label{A}
To help the reader in understanding how the general formula for the symmetrized product of chiral superfields emerges we 
summarize here the first few examples: 
\begin{eqnarray}
\Phi_1 \ast \Phi_2\vert_s & = & \Phi_1\Phi_2 -\text{det}C\bigg[\frac{\partial}{\partial
 \theta \theta}\Phi_1\bigg]\frac{\partial}{\partial
 \theta \theta}\Phi_2 \ ,
\\
\Phi_1 \ast \Phi_2 \ast \Phi_3\vert_s & =& \Phi_1 \Phi_2 \Phi_3 -\text{det}C\bigg[\Phi_1\frac{\partial}{\partial
\theta \theta}\Phi_2+ \frac{\partial}{\partial
\theta \theta}\Phi_1\Phi_2\bigg]\frac{\partial}{\partial
\theta \theta}\Phi_3\vert_s \ ,
\\
\Phi_1 \ast \cdots \ast \Phi_4 \vert_s & = & \Phi_1 \cdots \Phi_4 \nonumber 
\\ & & -\text{det}C\bigg[\Phi_1\Phi_2\frac{\partial}{\partial
\theta \theta}\Phi_3+\Phi_1\frac{\partial}{\partial
\theta \theta}\Phi_2\Phi_3 \nonumber \\ & & +\frac{\partial}{\partial
\theta \theta}\Phi_1\Phi_2\Phi_3\bigg]\frac{\partial}{\partial
\theta \theta}\Phi_4\vert_s \nonumber \\ & & +(\text{det}C)^2\bigg[\frac{\partial}{\partial
\theta \theta}\Phi_1\frac{\partial}{\partial
\theta \theta}\Phi_2\bigg]\frac{\partial}{\partial
\theta \theta}\Phi_3\frac{\partial}{\partial
\theta \theta}\Phi_4\vert_s \ ,
\\
\Phi_1 \ast \cdots \ast \Phi_5 \vert_s & = & \Phi_1 \cdots \Phi_5 \nonumber \\ & & -\text{det}C\bigg[\frac{\partial}{\partial
\theta \theta}\Phi_1\Phi_2\Phi_3\Phi_4+\Phi_1\frac{\partial}{\partial
\theta \theta}\Phi_2\Phi_3\Phi_4 \nonumber \\ & & +\Phi_1\Phi_2\frac{\partial}{\partial
\theta \theta}\Phi_3\Phi_4 +\Phi_1\Phi_2\Phi_3\frac{\partial}{\partial
\theta \theta}\Phi_4\bigg]\frac{\partial}{\partial
\theta \theta}\Phi_5\vert_s \nonumber \\ & & +(\text{det}C)^2\bigg[\frac{\partial}{\partial
\theta \theta}\Phi_1\frac{\partial}{\partial
\theta \theta}\Phi_2\Phi_3+\frac{\partial}{\partial
\theta \theta}\Phi_1\Phi_2\frac{\partial}{\partial
\theta \theta}\Phi_3 \nonumber \\ & & +\Phi_1\frac{\partial}{\partial
\theta \theta}\Phi_2\frac{\partial}{\partial
\theta \theta}\Phi_3\bigg]\frac{\partial}{\partial
\theta \theta}\Phi_4\frac{\partial}{\partial
\theta \theta}\Phi_5\vert_s \ .
\end{eqnarray}
It is then easy to guess the general formula:
\begin{equation}
  \Phi_1\ast\ldots\ast\Phi_n\vert_s = \Phi_1\cdot\ldots\cdot\Phi_n
  +\sum_{j=1}^{[\frac{n}{2}]}(-\det
  C)^jb_j(n)\prod_{k=n-j+1}^{n}\frac{\partial}{\partial\theta\theta}\Phi_k
  \vert_s \ ,
\end{equation}
where $b_j(n)$ is a polynomial in the fields of the form:
\begin{equation}
  b_j(n)=\sum_{i=0}^{n-2j}a_i(j,n)\Phi_0\cdots\Phi_i \ ,\qquad\qquad \Phi_0
  \equiv 1
\end{equation}
and the coefficients $a_i(j,n)$ are constructed as follows: Consider $j$
operators $\frac{\partial}{\partial\theta\theta}$. Then the
coefficient $a_i(j,n)$ is the sum of all possible terms of the form:
\begin{equation}
  \underbrace{\frac{\partial}{\partial\theta\theta}(\Phi_{i+1}\cdots\Phi_{l_1})
  \frac{\partial}{\partial\theta\theta}(\Phi_{l_1+1}\cdots\Phi_{l_2})
 \cdots\frac{\partial}{\partial\theta\theta}
  (\Phi_{l_{j-1}+1}\cdots\Phi_{n-j})}_{j\ \text{F-terms}} \ , 
\end{equation}
where the superfields are ordered according to
$\Phi_{i+1}\Phi_{i+2}\cdots\Phi_{n-j}$.  
\section{The K\"ahler potential }
In the main text we arrived at the expression
\begin{equation}
  \int d^4\theta\big(L^{i_1}\ast \ldots \ast L^{i_n}\ast \overline{L}^{j_1}\ast
  \overline{L}^{j_2}\big)\vert_{s\overline{s}} = \int d^4 \theta \big(
  L^{i_1}\ast\ldots\ast L^{i_n}\ast\overline{L}^{j_1}\big)
  \vert_{s\overline{s}} \overline{L}^{j_2} \ ,
\end{equation}
which we now express in terms of the components fields. In 
order to perform such a computation we now use once more (\ref{eq:16}). 
The analysis is then similar to the one for the
superpotential. This means that we will consider the antichiral field
$\overline{L}^{j_1}$ as, yet just another field and hence define:
\begin{eqnarray}\overline{L}^{j_1}\equiv L^{i_{n+1}} \ .\end{eqnarray} 
We now wish to compute the product:
\begin{equation}
  \label{eq:42}
  \big(L^{i_1}\ast\ldots\ast L^{i_n}\ast L^{i_{n+1}}\big) \vert_s \ ,
\end{equation}
where the symmetrization is in all of the $n+1$ indices. Since $L^{i_{n+1}}$
is no longer a chiral superfield the constraints
(\ref{eq:33}),(\ref{eq:34}) are no longer valid. However a slight
modification is still possible. First notice that
$L^{i_j}L^{i_k}L^{i_l}L^{i_m}=0$ with $j,k,l,m=1,\ldots ,n+1$. So we
can at maximum have three $L$'s standing next to each other in
$b_j(n)$. Also notice that the new constraints are in force: 
\begin{equation}
  \label{eq:43}
  L^{i_j}L^{i_k}L^{i_l}\frac{\partial}{\partial\theta\theta}
  L^{i_m}L^{i_n}L^{i_r}=0 \ ,
\end{equation}

\begin{equation}
  \label{eq:44}
  L^{i_j}L^{i_k}L^{i_l}\frac{\partial}{\partial\theta\theta}L^{i_m}L^{i_n}
  \vert_s = L^{i_j}L^{i_k}\frac{\partial}{\partial\theta\theta}
  L^{i_l}L^{i_m}L^{i_n}\vert_s =0 \ ,
\end{equation}

\begin{equation}
  \label{eq:45}
  L^{i_j}L^{i_k}\frac{\partial}{\partial\theta\theta}L^{i_l}L^{i_m}
  \frac{\partial}{\partial\theta\theta}L^{i_n}L^{i_r}\vert_s =0 \ ,
\end{equation}
where (\ref{eq:44}) and (\ref{eq:45}) follow from the spin-statistics
theorem. We see that the number of fields $n+1-j$ can be no larger than
the number of operators $j$ plus three: $n+1-j \leq j+3$. Hence in
order to obtain non-vanishing contributions we must have: 
\begin{equation}
  \label{eq:46}
  \frac{n}{2}-1 \leq j \leq \bigg[\frac{n+1}{2}\bigg] \ .
\end{equation}
So if $n$ is even we have $j=\frac{n}{2}-1$ and $j=\frac{n}{2}$. If
$n$ is odd we have $j=\frac{n-1}{2}$ and $j=\frac{n+1}{2}$. For $n=2m$
we get:
\begin{eqnarray}
  \label{eq:47}
  L^{i_1}\ast\ldots\ast L^{i_{2m+1}}\vert_s & = & (-\det
  C)^{m-1} \bigg( L^{i_1}L^{i_2}L^{i_3}\frac{\partial}
  {\partial\theta\theta}L^{i_4}\cdots \frac{\partial}{\partial\theta
  \theta}L^{i_{2m+1}}
  \nonumber
  \\
   & & +(m-1)L^{i_1}L^{i_2}
  \frac{\partial}{\partial\theta\theta}
  L^{i_3}L^{i_4}\frac{\partial}{\partial\theta\theta} L^{i_5}\cdots
  \frac{\partial}{\partial\theta\theta}L^{i_{2m+1}}
  \nonumber
  \\
   & & +\frac{1}{2}(m-2)(m-1)L^{i_1} 
  \frac{\partial}{\partial\theta\theta}L^{i_2}L^{i_3}
  \frac{\partial}{\partial\theta\theta}L^{i_4}L^{i_5}
  \nonumber
  \\
  & & \times \frac{\partial}{\partial\theta\theta}L^{i_6}\cdots
  \frac{\partial}{\partial\theta\theta}L^{i_{2m+1}}
  \nonumber
  \\
   & &
  +(m-1)L^{i_1}\frac{\partial}{\partial\theta\theta}L^{i_2}
  L^{i_3}L^{i_4}\frac{\partial}{\partial\theta\theta}L^{i_5}\cdots
  \frac{\partial}{\partial\theta\theta}L^{i_{2m+1}}\bigg)\vert_s
  \nonumber
  \\
   & & +(-\det
  C)^m\bigg(L^{i_1}\frac{\partial}{\partial\theta
  \theta}L^{i_2}\cdots \frac{\partial}{\partial\theta\theta}
  L^{i_{2m+1}}\bigg)\vert_s
  \nonumber
  \\
  & & +\textrm{Terms containing no}\ \theta \textrm{'s} \ .
\end{eqnarray}
The factor in the third line is determined as follows. Assume that there exist
$j=m-1$ operators. Then we can arrange the $j+2$ fields in:
\begin{equation}
  \label{eq:48}
  (j-1)+(j-2)+\ldots + 1  =  \sum_{s=1}^{j-1}(j-s) 
   =  \frac{1}{2}j(j-1) \ ,
\end{equation}
different ways, where the fields are ordered in the products. For $n=2m+1$ we obtain:
\begin{eqnarray}
  \label{eq:49}
  L^{i_1}\ast\cdots\ast L^{i_{2m+2}}\vert_s & = & (-\det C)^m\bigg(
  L^{i_1}L^{i_2}\frac{\partial}{\partial\theta\theta}L^{i_3}\cdots 
  \frac{\partial}{\partial\theta\theta}L^{i_{2m+2}} 
  \nonumber
  \\
  & & +m L^{i_1} \frac{\partial}{\partial\theta\theta} L^{i_2} L^{i_3}
  \frac{\partial}{\partial\theta\theta} L^{i_4}\cdots \frac{\partial}
  {\partial\theta\theta}L^{i_{2m+2}}\bigg)\vert_s
  \nonumber
  \\
  & & +\textrm{Terms containing no}\ \theta \textrm{'s} \ .
\end{eqnarray}
It is now time to recall that $L^{i_{n+1}}=\overline{L}^{j_1}$. If we
think a bit ahead we see that in the end we have to multiply by 
$\overline{L}^{j_2}$ and then integrate over the superspace. The terms
containing no $\theta$'s must have at least one
$\overline{\theta}$. The only term of $\overline{L}^{j_2}$ containing
a $\theta\theta$ (required from superspace integration) also 
contains a $\overline{\theta\theta}$ and 
therefore the terms containing no $\theta$'s of (\ref{eq:47}) and 
(\ref{eq:49}) vanish in the end. Also we cannot have
$\overline{L}^{j_1}$ standing alone with an operator acting on
it, simply because it will only contain a $\overline{\theta\theta}$
and therefore vanishes when multiplied with
$\overline{L}^{j_2}$. Therefore when symmetrizing in $L^{i_{n+1}}=
\overline{L}^{j_1}$ we obtain for $n=2m$: 
\begin{eqnarray}
  \label{eq:50}
  L^{i_1}\ast\cdots\ast
  L^{i_{2m}}\ast\overline{L}^{j_1}\vert_{s\overline{s}}
  & = & \frac{(-\det C)^{m-1}}{2m+1}\bigg(
  3\overline{L}^{j_1}L^{i_1}L^{i_2}
  \frac{\partial}{\partial\theta\theta}L^{i_3}\cdots\frac{\partial}{\partial
  \theta\theta}L^{i_{2m}}
  \nonumber
  \\
  & &
  +2(m-1)\overline{L}^{j_1}L^{i_1}\frac{\partial}{\partial\theta\theta}
  L^{i_2}L^{i_3}\frac{\partial}{\partial\theta\theta}L^{i_4}\cdots
  \frac{\partial}{\partial\theta\theta}L^{i_{2m}}
  \nonumber
  \\
  & & +2(m-1)L^{i_1}L^{i_2}\frac{\partial}{\partial\theta\theta}\overline{L}
  ^{j_1}L^{i_3}\frac{\partial}{\partial\theta\theta}L^{i_4}\cdots
  \frac{\partial}{\partial\theta\theta}L^{i_{2m}}
  \nonumber
  \\
  & & +2(m-2)(m-1)L^{i_1}\frac{\partial}{\partial\theta\theta}\overline{L}
  ^{j_1}L^{i_2}\frac{\partial}{\partial\theta\theta}L^{i_3}L^{i_4}
  \nonumber
  \\
  & & \times \frac{\partial}{\partial\theta\theta}L^{i_5}\cdots
  \frac{\partial}{\partial\theta\theta}L^{i_{2m}}
  \nonumber
  \\
  & & +3(m-1)L^{i_1}\frac{\partial}{\partial\theta\theta}\overline{L}^{j_1}L^{
  i_2}L^{i_3}
  \nonumber
  \\
  & & \times \frac{\partial}{\partial\theta\theta}L^{i_4}\cdots 
  \frac{\partial}{\partial\theta\theta}L^{i_{2m}}\bigg)\vert_s
  \nonumber
  \\
  & & +\frac{(-\det
  C)^m}{2m+1}\bigg(\overline{L}^{j_1}\frac{\partial}{\partial\theta\theta}
  L^{i_1}\cdots \frac{\partial}{\partial\theta\theta}
  L^{i_{2m}}\bigg)\vert_s
  \nonumber
  \\
  & & +\textrm{Terms not contributing to}\ \mathcal{L} \ .
\end{eqnarray}

{}For $n=2m+1$ we get:
\begin{eqnarray}
  \label{eq:51}
  L^{i_1}\ast\cdots\ast
  L^{i_{2m+1}}\ast\overline{L}^{j_1}\vert_{s\overline{s}} & = & \frac
  {(-\det C)^m}{2m+2}\bigg( 2\overline{L}^{j_1}
  L^{i_1}\frac{\partial}{\partial\theta\theta}L^{i_2}\cdots
  \frac{\partial}{\partial\theta\theta}L^{i_{2m+1}}
  \nonumber
  \\
  & & +m\overline{L}^{j_1}\frac{\partial}{\partial\theta\theta}
  L^{i_1}L^{i_2}
  \frac{\partial}{\partial\theta\theta}L^{i_3}\cdots
  \frac{\partial}{\partial\theta\theta}L^{i_{2m+1}}
  \nonumber
  \\
  & &
  +2mL^{i_1}\frac{\partial}{\partial\theta\theta}\overline{L}^{j_1}L^{i_2}
  \frac{\partial}{\partial\theta\theta}L^{i_3}\cdots 
  \frac{\partial}{\partial\theta\theta}L^{i_{2m+1}}\bigg)\vert_s
  \nonumber
  \\
  & & +\textrm{Terms not contributing to}\ \mathcal{L} \ .
\end{eqnarray}
It is now time that we write it out in components. If we remember that
every time a term contains three spinors or more it vanishes after
symmetrization and if we use the fact that the expansion coefficients
are symmetric in their indices we obtain for $n$ even:
\begin{displaymath}
  \int d^4 \theta \big(L^{i_1}\ast\cdots\ast L^{i_{2m}}\ast
  \overline{L}^{j_1}\ast\overline{L}^{j_2}\big) \vert_{s\overline{s}} 
  K_{,i_1\cdots i_{2m} j_1 j_2} 
\end{displaymath}
\begin{displaymath}
  =\bigg((-\det
  C)^{m-1}m\overline{\psi}^{j_1}\overline{\psi}^{j_2}\psi^{i_1}\psi^
  {i_2}F^{i_3}\cdots F^{i_{2m}}
\end{displaymath}
\begin{equation}
  \label{eq:52}
  +(-\det C)^m \frac{2}{2m+1}\partial^{\mu}
  \overline{A}^{j_1}\partial_{\mu}\overline{A}^{j_2}F^{i_1}\cdots
  F^{i_{2m}}\bigg)
   K_{,i_1\cdots i_{2m} j_1 j_2} \ ,
\end{equation}
and for $n$ odd:
\begin{displaymath}
  \int d^4 \theta \big(L^{i_1}\ast\cdots\ast L^{i_{2m+1}}\ast
  \overline{L}^{j_1}\ast\overline{L}^{j_2}\big) \vert_{s\overline{s}} 
  K_{,i_1\cdots i_{2m+1} j_1 j_2} 
\end{displaymath}
\begin{displaymath}
  =(-\det
  C)^m\bigg(-2i\psi^{i_1}\sigma^{\mu}\overline{\psi}^{j_1}\partial_{\mu}
  \overline{A}^{j_2}F^{i_2}\cdots F^{i_{2m+1}}
\end{displaymath}
\begin{equation}
  \label{eq:53}
  -\overline{\psi}^{j_1}\overline{\psi}^{j_2}F^{i_1}\cdots
   F^{i_{2m+1}}\bigg) K_{,i_1\cdots i_{2m+1} j_1 j_2} \ .
\end{equation}
The contribution to the Lagrangian therefore is:
\begin{displaymath}
  \int d^4 \theta \sum_{n=1}^{\infty}\frac{1}{n!2}\big(L^{i_1}\ast\cdots 
  \ast L^{i_n}\ast
  \overline{L}^{j_1}\ast\overline{L}^{j_2}\big)\vert_{s\overline
  {s}} K_{,i_1\cdots i_{n} j_1 j_2} 
\end{displaymath}
\begin{displaymath}
  =\int d^4 \theta \bigg\{\frac{1}{2}L^{i_1}\overline{L}^{j_1}
\overline{L}^{j_2}K_{,i_1 j_1 j_2} 
+\frac{1}{4}L^{i_1}L^{i_2}
  \overline{L}^{j_1}
  \overline{L}^{j_2}K_{,i_1 i_2 j_1 j_2}\bigg\} 
\end{displaymath}
\begin{displaymath}
  +\sum_{n=2}^{\infty}\frac{(-\det
  C)^{n-1}}{(2n)!}\frac{n}{2}\overline{\psi}^{j_1}\overline{\psi}^{j_2}\psi^{i_1}
  \psi^{i_2}F^{i_3}\cdots F^{i_{2n}}K_{,i_1\cdots i_{2n} j_1 j_2}
\end{displaymath}
\begin{displaymath}
  +\sum_{n=1}^{\infty}\frac{(-\det
  C)^n}{(2n+1)!}\partial^{\mu}\overline{A}^{j_1}
  \partial_{\mu}\overline{A}^{j_2}
  F^{i_1}\cdots F^{i_{2n}} K_{,i_1\cdots i_{2n} j_1 j_2} 
\end{displaymath}
\begin{equation}
  \label{eq:54?}
  -\sum_{n=1}^{\infty}\frac{(-\det
   C)^n}{(2n+1)!}\bigg\{i\psi^{i_1}\sigma^{\mu}
  \overline{\psi}^{j_1}\partial_{\mu}\overline{A}^{j_2}+\frac{1}{2}\overline{\psi}^{j_1}
  \overline{\psi}^{j_2}F^{i_1}\bigg\}F^{i_2}\cdots F^{i_{2n+1}}
  K_{,i_1\cdots i_{2n+1} j_1 j_2} \ .
\end{equation}
 
Finally together with (\ref{eq:37}), (\ref{eq:18}),(\ref{eq:54}) 
 as well as (\ref{eq:38}) the entire Lagrangian can be constructed.

\section{Single Field Limit}
 {}For the reader's convenience we specialize the theory to the case of a single field which was first 
 investigated in \cite{Azorkina:2005mx}:
\begin{eqnarray}
  \label{eq:56}
  \mathcal{L} & = & \int d^4 \theta K(\Phi,\overline{\Phi}) + \int
  d^2 \theta P(\Phi) + \int
  d^2\overline{\theta}\overline{P}(\overline{\Phi})
  \nonumber
  \\
  & & -\sum_{n=2}^{\infty}\frac{(-\det C)^{n-1}}{(2n)!}n\psi\psi
  F^{2n-2}\overline{F}
  \frac{\partial^{2n+1}}{\partial\overline{A}\partial
  A^{2n}} K(A,\overline{A})
  \nonumber
  \\
  & & +\sum_{n=1}^{\infty} \frac{(-\det C)^n}{(2n)!}F^{2n}
   \partial^2 \overline{A} \frac{\partial^{2n+1}}{\partial\overline{A}\partial
  A^{2n}} K(A,\overline{A})
  \nonumber
  \\
  & & +\sum_{n=1}^{\infty}\frac{(-\det
  C)^n}{(2n+1)!}F^{2n+1}\overline{F}\frac{\partial^{2n+2}}
  {\partial\overline{A}\partial A^{2n+1}} K(A,\overline{A})
  \nonumber
  \\
   & & +\sum_{n=1}^{\infty}\frac{(-\det
  C)^n}{(2n+1)!}i\partial_{\mu}\overline{\psi}\overline{\sigma}^{\mu}\psi
  F^{2n} \frac{\partial^{2n+2}}
  {\partial\overline{A}\partial A^{2n+1}} K(A,\overline{A})
  \nonumber
  \\
  & & -\sum_{n=1}^{\infty}\frac{(-\det
  C)^n}{(2n+1)!}n\psi\psi F^{2n-1}\partial^2 \overline{A} 
  \frac{\partial^{2n+2}}
  {\partial\overline{A}\partial A^{2n+1}} K(A,\overline{A})
  \nonumber
  \\
  & & +\sum_{n=2}^{\infty}\frac{(-\det
  C)^{n-1}}{(2n)!}\frac{n}{2}\overline{\psi}\overline{\psi}\psi\psi F^{2n-2} 
  \frac{\partial^{2n+2}}
  {\partial\overline{A}^2\partial A^{2n}} K(A,\overline{A})
  \nonumber
  \\
  & & +\sum_{n=1}^{\infty}\frac{(-\det
  C)^n}{(2n+1)!}F^{2n}\partial^{\mu}\overline{A}\partial_{\mu}
  \overline{A}\frac{\partial^{2n+2}}
  {\partial\overline{A}^2\partial A^{2n}} K(A,\overline{A})
  \nonumber
  \\
  & & -\sum_{n=1}^{\infty}\frac{(-\det
  C)^n}{(2n+1)!}i\psi\sigma^{\mu} \overline{\psi}\partial_{\mu}\overline{A}
  F^{2n}\frac{\partial^{2n+3}}
  {\partial\overline{A}^2\partial A^{2n+1}} K(A,\overline{A})
  \nonumber
  \\
  & & -\sum_{n=1}^{\infty}\frac{(-\det
  C)^n}{(2n+1)!}\frac{1}{2}\overline{\psi}\overline{\psi}F^{2n+1} \frac{\partial^{2n+3}}
  {\partial\overline{A}^2\partial A^{2n+1}} K(A,\overline{A})
  \nonumber
  \\
  & & +\sum_{n=1}^{\infty}\frac{(-\det C)^n}{(2n+1)!}F^{2n+1}\frac
  {\partial^{2n+1}}{\partial A^{2n+1}}P(A)
  \nonumber
  \\
  & & -\sum_{n=2}^{\infty}\frac{(-\det C)^{n-1}}{(2n)!}n\psi\psi
  F^{2n-2} \frac{\partial^{2n}}{\partial A^{2n}}P(A) \ .
\end{eqnarray}

\end{document}